\documentclass[a4paper,12pt]{article}
\usepackage{epsfig}
\usepackage{axodraw}
\usepackage{amsmath,amssymb}

\newcommand\email[1]{{\it #1}}
\newcommand\institution[1]{{\it #1}}
\newcommand\address[1]{{\it #1}}
\begin{document}
\mathchardef\vm="117
\mathchardef\um="11D
\mathchardef\E="245
\mathchardef\Mom="250
\def\z1{z{}'}
\def\t1{t{}'}
\def\u1{u{}'}
\def\s1{s{}'}
\def\m2{m^2}
\def\d12{\frac{1}{2}}
\def\lfr#1#2{\ln{\frac{#1}{#2}}}

\def\ub{\bar{\um}}
\def\Jint#1{\mathcal{J} ( {#1} )}
\def\Iint#1#2{\int\limits_{0}^{\ub}{\Jint{#1}} {#2} {d \um}}
\def\Iintl#1#2{\int\limits_{2 m \lambda}^{\ub}{\Jint{#1}} {#2} {d \um}}

\def\bracket#1{\left({#1}\right)}
\def\bra#1{\bracket{#1}}
\def\spr#1#2{\, #1 \! \cdot \! #2 \,}
\def\ddx#1{\frac{1}{#1}}

\def\umax{\um_{m}}
\def\sminumax{s-\um_{m}}
\def\Fu#1{\mathcal{F}\bra{#1}}
\def\spens#1{\Phi\bra{#1}}
\title{Search for effects beyond Standard Model in photon scatterings and in nonminimal gauge theories on linear colliders of new generation}
\author{T. V. Shishkina}
\date{}
\maketitle
\vspace{-5mm}
\begin{center}
\institution{Department of Theoretical Physics, Belarusian State University\\}
\address{Nezavisimosti av. 4, Minsk 220030, Belarus\\}
\email{e-mail: shishkina@bsu.by}
\end{center}
\vspace{5mm}
%

\abstract{The main possibilities of investigation of
leptons and bosons production
in interaction of polarized photons are considered.
The usage of
$\gamma \gamma \to f \bar{f} [+\gamma]$ reactions for the luminosity measurement
on linear photon collider is analyzed.
The achievable precision of the luminosity measuring is considered and calculated.
The first-order QED correction to $\gamma\gamma\to l \bar{l}$
scattering is analyzed.
All possible polarization
states of interacting particles are investigated.
For the detection of deviations from SM predictions
at linear $\gamma\gamma$ colliders with center of mass energies running to $1\, TeV$
the influence of three possible
anomalous couplings on the cross sections of $W^+W^-$
productions has been investigated.
The significant discrimination between various anomalous contributions is discovered.
The main contribution of high order electroweak effects is considered.}

\section{Introduction}
The Standard Model (SM) has possibility to describe all
experimental data up to now
with typical precision around
one per mil. Nevertheless the Model is
not the final theory valid up
to very high scales and
at linear collider that can run at centre of mass energies around 1 TeV
one can hope to see finally deviations in precision measurements occur typically for two reasons.

If the new physics occurs in loop diagrams
their effect is usually suppressed by a loop factor $\alpha \slash {4 \pi}$
and very high precision is required to see it.
If the new physics is already
on the Born level but at very high masses
the effects are suppressed by propagator factor
${s}/({s-m^2_{N_P}-\imath m_{N_P} \Gamma})$
so that is important to work at the highest possible energies.

Linear lepton colliders will provide the opportunity to
investigate photon collisions at energies and luminosities close
to these in $e^{+} e^{-}$ collisions \cite{gg_proposal}.

The possibility to transform the future linear
$e^{+}e^{-}$-colliders into the $\gamma \gamma$ and $\gamma e$
-colliders with approximately the same energies and luminosities
was shown. The basic $e^+ e^-$-colliders can be transformed into
the $e \gamma$- or $\gamma \gamma$-colliders. The intense $\gamma$
-beams for photon colliders are suggested to be obtained by
Compton scattering of laser lights which is focused on electrons
beams of basic $e^+ e^-$-accelerators.

The electron and photon linear colliders of next generation will attack
unexplored higher energy region where new behaviour can turn up.
In this area the photon colliders have a number of advantages.

-- The first of the above advantages is connected with
the better signal/background ratio at both $e^+e^-$- and
$e\gamma\slash\gamma\gamma$-colliders in comparison with hadron ones.

-- The production cross sections at photon colliders are usually larger
than those at electron colliders.

-- The photon colliders permit to investigate both of the problems
of new physics and those ones of "classical" hadron physics and QCD.

Compare of above mentioned electron and photon colliders.

1. In the scheme considered the maximal photon energy is slightly less than
electron energy $E$.

To increase the maximal photon energy one can use the laser with
the largest frequency. It seems also useful to do photon spectrum
more monochromatic. However with such energy growth the new
phenomenon takes place which destroy the obtained photon beams.
The high energy photons disappear from the beam due to their
collisions with laser ones producing $e^+ e^-$ -pairs.

2. The $e \gamma$ and $\gamma \gamma$ luminosities can be the same as basic
$e^+ e^-$ luminosity or even larger (for instance for $\gamma \gamma$
collisions).

3. It seems to be an important advantage of the electron beams that they
are the monochromatic ones.
It isn't correct.

Really the production of the heavy particles in electron colliders can be
represented as two-step process. At the first step an electron
emits photons (it is standard bremsstrahlung -- initial state
radiation). After that the electrons with the lower energies
collide and produce the heavy particles.
Secondly, the electron spectrum is smoothed due to bremsstrahlung.
This spectrum varies during bunch collision.

4. The photon spectrum is nonmonochromatic.
Its effective form depends on the conditions of the conversion.
Besides the collisions of electron with a few laser photons simultaneously
result in high energy tail of spectrum (nonlinear QED effect).
On the other hand due to angular spread of photons the effective form of their
spectrum varies with the distance between conversion and collision points.

5.
Only with using of detailed
data on momenta of particles observed one can restore
the real energy dependencies of cross sections.
The determination of cross section averaged over the above wide spectra seems
to be useful for very preliminary estimations only.

At the colliders discussed the data processing should be performed
with equation of the form:
\begin{eqnarray}
\int \frac{\partial^2 L(E_1,
E_2)}{\partial E_1 \partial E_2} \cdot \sigma(W^2)
{\mid}_{W^2=4E_1E_2} \cdot dE_1 dE_2.
\end{eqnarray}
Therefore the special
measurements of the spectral luminosity $dL(E_1, E_2)$ (i. e. the
distribution of luminosity in $W$ and in the rapidity of produced
system) are necessary. The preliminary estimations shows that one
could use for this aim the Bhabha scattering for electron
colliders, the Bethe-Heither $e \gamma \to e \mu^+ \mu^-$
processes for $e \gamma$ - colliders, $\gamma \gamma\to
\mu^+\mu^-\mu^+\mu^-$ process for $\gamma\gamma$ -colliders.

6. In the $e^+e^-$ -colliders the region of small angles closed
for the observations.
The small angle region will be open for investigation at
$\gamma\gamma$ and $\gamma e$- colliders.

7.
The degree of photon polarization correlates with its energy. The
polarization of hard photons can be calculated: the special
measurements for soft tail are needed. The same problem for
electrons is due to the variation of their polarizations induced
by bremsstrahlung.

8. In the $e^+e^-$ -collisions in the most cases the states $J=1$ are produces.
Therefore, the $e^+e^-$ -colliders are suitable for investigation of neutral
vector bosons.

At the $\gamma\gamma$ -colliders all the partial waves are produced.
The set physical processes which can be investigated at the
$\gamma\gamma$ -colliders is richer than that in the $e^+e^-$-colliders.

9. The production cross section at $\gamma\gamma$ collisions are usually
larger than those ones at $e^+e^-$-collisions and they are decreased
slowly with the energy. It is the source of the additive advantage of
$\gamma \gamma$ colliders because the detailed investigation of many
reactions and particles is preferable for above the threshold.

10.
There is no need in the positron beams for the $\gamma e$ and $\gamma \gamma$
colliders. It is sufficient to have as a base the
$e^-e^-$ collider only.

So it is exclusively important task
to use possibilities of $\gamma\gamma$-colliders to realize the
experiments of the next generation.

If a light Higgs exists one of the main tasks of a photon collider will be the measurement
of the partial width $\Gamma (H \to \gamma\gamma)$.
Not to be limited by the error from luminosity determination
the luminosity of the collider at the energy
of the Higgs mass has to be known with a precision of around $1 \%$.

To produce scalar Higgses the total angular momentum of the two photons
has to be $J\!=\!0$.
In this case the cross section
$\gamma\gamma \to l^{+} l^{-}$ is suppressed by factor $m^2_{l} \slash s$
and thus not usable for luminosity determination.

In the SM the couplings of the
gauge bosons and fermions are constrained
by the requirements of gauge symmetry.
In the electroweak sector this
leads to trilinear and quartic
interactions between the gauge bosons
with completely specified couplings.

The trilinear and quartic
gauge boson couplings probe different
aspects of the weak interactions.
The trilinear couplings directly test the
non-Abelian gauge structure, and possible
deviations from the SM forms have been
extensively studied. In contrast,
the quartic couplings can be
regarded as a more direct way
of consideration of electroweak symmetry
breaking or, more generally, on
new physics which couples to electroweak
bosons.

In this respect it is quite possible
that the quartic couplings deviate
from their SM values while the
triple gauge vertices do not.
For example, if the mechanism
for electroweak symmetry breaking
doesn't reveal itself through the discovery
of new particles such as the
Higgs boson, supersymmetric particles
or technipions it is possible
that anomalous quartic couplings
could provide the first evidence
of new physics in this sector
of electroweak theory.

The production of several vector bosons
is the best place to search directly
for any anomalous behaviour of
triple and quartic couplings.

By using of transforming a linear $e^{+}e^{-}$ collider in a
$\gamma\gamma$ collider, one can obtain very energetic photons
from an electron or positron beams. Such machines as ILC which
will reach a center of mass energy $\sim 1000 GeV$ with high
luminosity ($\sim 10^{33} cm^{-2} s^{-1}$) will be able to study
multiple vector boson production with high statistics.

For obvious kinematic reasons,
processes where at least one
of the gauge bosons is a photon
have the largest cross sections.

So
the photon linear colliders have the great physical potential \cite{gg_other}
(Higgs and SUSY particles searching,
study of anomalous gauge boson couplings
and hadronic structure of photons etc.).
Performing of this set of investigations
requires a fine measurement of the luminosity of photon beams.
For this purpose some of the well-known
and precisely calculated reactions
(see, for example, $\gamma\gamma\to 2 f, 4 f$ \cite{gg_llg_excl,gg_ll_corr,gg_2f_denner,gg_4l_moretti,gg_4l_kapusta,gg_4l_minsk}) are traditionally used.

It was shown that
it is convenient to use the events of $\gamma\gamma\to l^+l^-$ process
for measuring the luminosity of the $J\!=\!2$-beams
($J$ is the total angular momentum of initial photon couple).
Here $l$ is the unpolarized light lepton ($e$ or $\mu$).
It is the dominating QED process on $J\!=\!2$ beams
and its events are easily detected.

The difficulties appear in the calibration of
photon beams of similar helicity
(the total helicity of $\gamma \gamma$-system $J\!=\!0$)
since the small magnitude of cross sections
of the most QED processes.
For example, the leading term of cross section of $\gamma \gamma \to l \bar{l}$
scattering on $J\!=\!0$-beams is of order $\alpha \slash \pi$ ($\approx 0.002$).

The exclusive reaction $\gamma\gamma\to l^+l^-\gamma$
provides the unique opportunity to measure the luminosity of $J\!=\!0$ beams
on a linear photon collider.

One of the main purposes of the linear photon collider
is the $s$-channel of the Higgs boson production
at energies about $\sqrt{s}=120 GeV$.
That is the reason of using this value of c.m.s. energy in our analysis.

\section{ Two lepton production with photon in $\gamma\gamma$-colli\-sions }

The two various helicity
configuration of the $\gamma \gamma$-system leads to the different spectra
of final particles
and requires the two mechanisms of beam calibration.
We have analyzed \cite{gg_llg_excl} the behaviour of the
$\gamma\gamma\to l^+l^-\gamma$ reaction
on beams with various helicities
as a function of the parameters of detectors,
and performed the detail comparison of cross section on $\gamma^+\gamma^+$-
($J=0$) and $\gamma^+\gamma^-$-beams  ($J=2$).
Since experimental beams are partially polarized
the ratio of cross sections of $\gamma \gamma \to l^+ l^- \gamma$ scattering on $J\!=\!0$ to $J\!=\!2$-beams
should be high
for the effective luminosity measurement.
We have outlined the conditions that greatly restrict the
observation of the process on $J=2$ beams, remaining the $J=0$ cross
section almost unchanged.

Finally we estimate the precision of luminosity measurement.

Consider the process
\begin{eqnarray}\label{proc3}
\gamma(p_1, \lambda_1)+\gamma(p_2, \lambda_2) \to f(p_1{}', e_1{}') + \bar{f}(p_2{}', e_2{}') + \gamma(p_3, \lambda_3),
\end{eqnarray}
where $\lambda_{i}$ and $e_{i}{}'$ are the photon and the fermion helicities.

We denote the c.m.s. energy squared by $s = {\left(p_1+p_2\right)}^2 = 2 \spr{p_1}{p_2},$
the final-state photon energy by $w$.
For the differential cross section the normalized
final-state photon energy (c.m.s. is used) $x=w\slash\sqrt{s}$ is introduced.
The differential cross section ${d\sigma}\slash{d x}$
appears to be the energy spectrum of
final-state photons.

The matrix elements are obtained using two methods: the massless
helicity amplitudes \cite{ha} for the fast estimations and the exact
covariant analysis \cite{shum,peaking} including finite fermion mass. Since
final-state polarizations will not be measured we have summarized
over all final particles helicities. The integration over the
phase space of final particles is performed numerically using the
Monte-Carlo method \cite{mc}.

The calculations have been performed for various experimental restrictions on
the parameters of final particles.
Events are not detected if energies and angles are below the corresponding threshold values.
The considering restrictions on the phase-space of final particles (the cuts) are denotes as follows:

$\bullet$ Minimum final-state photon energy: $\omega_{cut}$,

$\bullet$ Minimum fermions energy: $E_{f,cut}$,

$\bullet$ Minimum angle between any final and any initial particles (polar angle cut): $\Theta_{cut}$,

$\bullet$ Minimum angle between any pair of final particles: $\varphi_{cut}$.

\begin{figure}[!ht]
\leavevmode
\begin{minipage}[b]{.5\linewidth}
\centering
\includegraphics[width=\linewidth, height=2.6in, angle=0]{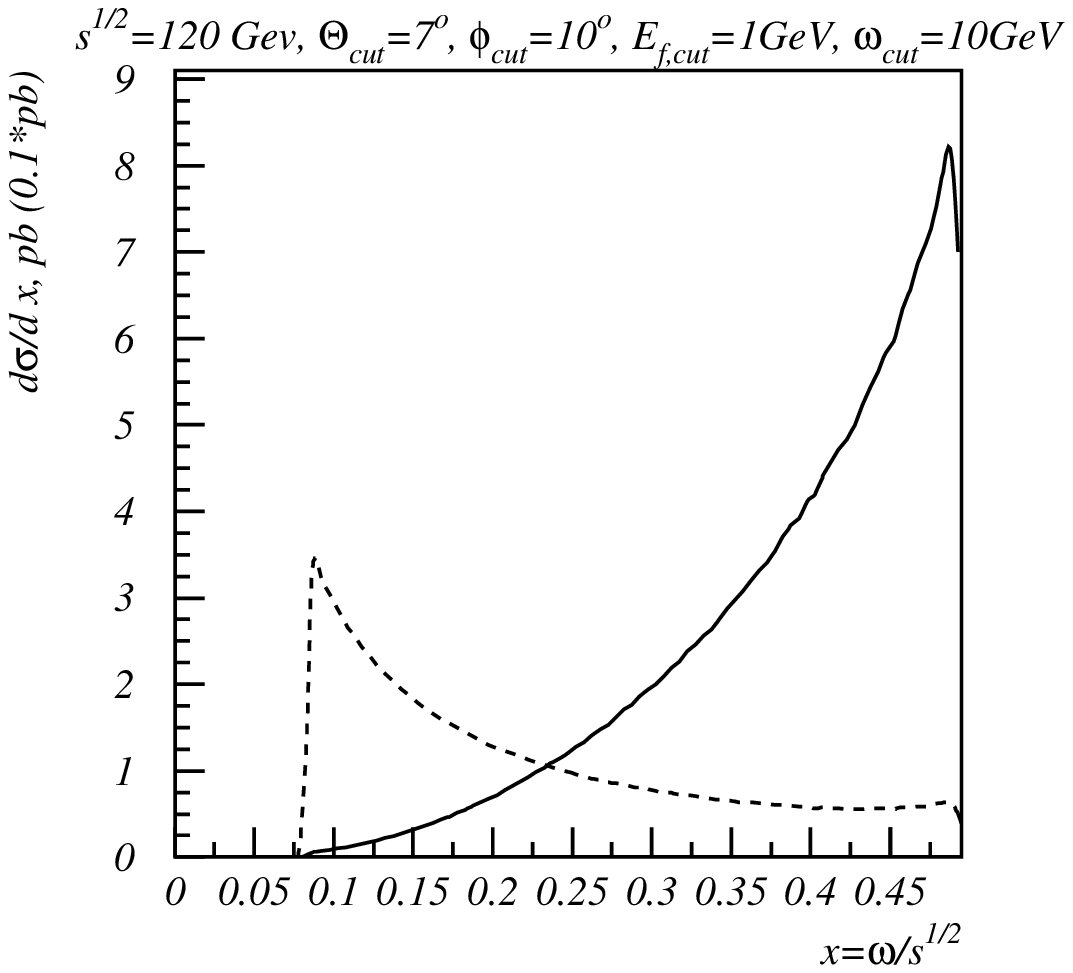}
\end{minipage}\hfill
\begin{minipage}[b]{.5\linewidth}
\centering
\includegraphics[width=\linewidth, height=2.6in, angle=0]{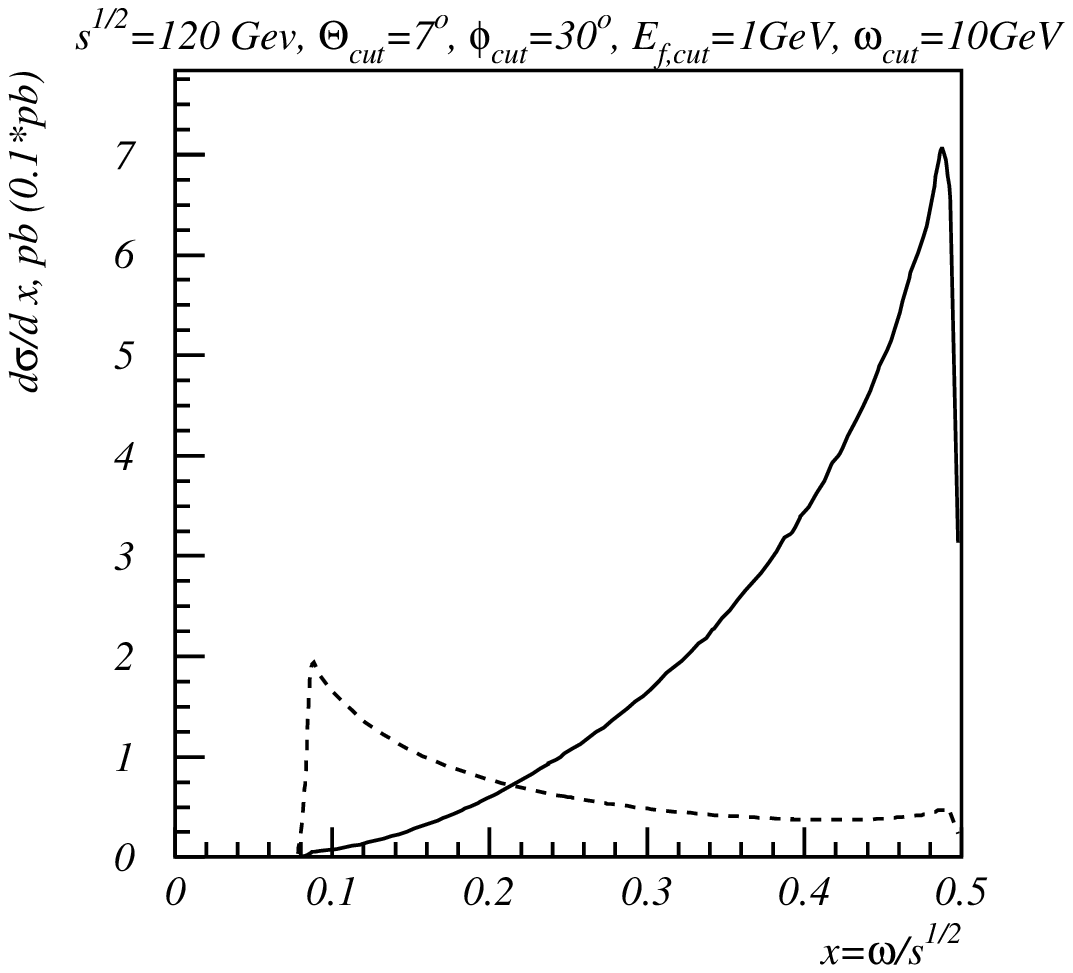}
\end{minipage}
\vspace{-30pt}
\caption{
Final-state photon energy spectrum for $J\!=\!0$ (solid) and $(J\!=\!2)*0.1$ (dotted)
at $\sqrt{s}=120GeV$ and cuts:
$\Theta_{min} \!=\! 7^o$, $\varphi_{min} \!=\! 10^o$ (left) and $\varphi_{min} \!=\! 30^o$ (right),
$E_{f, min} \!=\! 1 GeV$, $\omega_{min} \!=\! 10 GeV$.
}\label{diffs_llg}
\end{figure}

Consider the energy spectrum of final photons.
In Fig. \ref{diffs_llg} the spectra for $J\!=\!0$ and $(J\!=\!2)$ are presented
(the $(J\!=\!2)$-cross section is scaled on factor $0.1$ for the convenience).
The differential cross section
${d \sigma}\slash{d x}$ on $J=2$ beams
decreases while one on $J=0$ beams raises with increasing of the
final-state photon energy.
This leads to the conclusion that if one increases the threshold
on $w$,
the process on $J\!=\!2$ beams will be greatly restricted,
but the rate of $\!J=0\!$ events remains almost unchanged.

The ratio of events on $J\!=\!0$ and $J\!=\!2$ beams
strongly depends on the experimental cuts.
We obtained the region (the configuration of cuts)
where the processes on the both $J\!=\!0$ and $J\!=\!2$ beams
have the cross sections close by each other.
That is the region of small polar angle cut,
high collinear angle cut
as well as high minimal energy of final-state photons.
At these parameters the total cross sections
of $ \gamma\gamma\to f \bar{f}\gamma$ in
experiments using $\gamma^+\gamma^+$- and $\gamma^+\gamma^-$- beams
appear to be the same order of magnitude.

\begin{figure}[!ht]
\leavevmode
\begin{minipage}[h]{.5\linewidth}
\centering
\includegraphics[width=\linewidth, height=2.4in, angle=0]{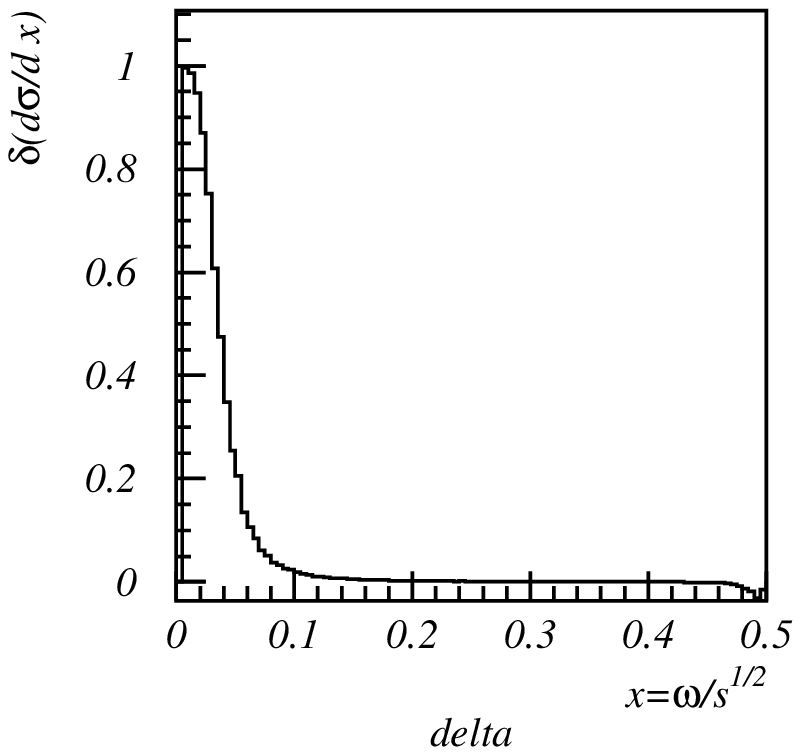}
\end{minipage}\hfill
\begin{minipage}[h]{.5\linewidth}
\centering
\includegraphics[width=\linewidth, height=2.4in, angle=0]{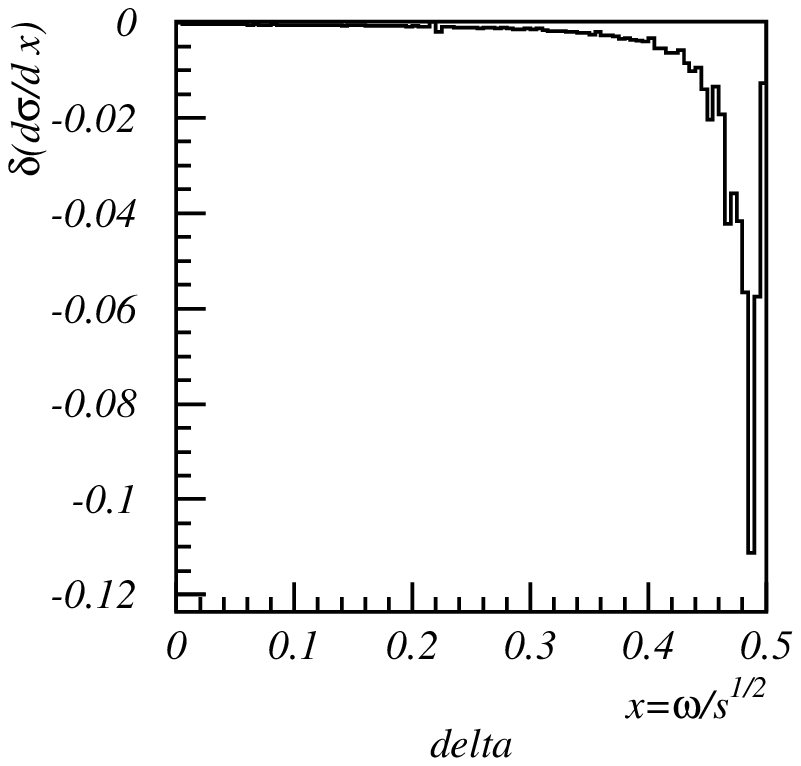}
\end{minipage}
\vspace{-10pt}
\caption{
The relative mass contribution to
energy spectra of final photon
for $J\!=\!0$ (left) and $J\!=\!2$ (right) beams
($w_{cut}\!=\!1GeV$, $\E_{cut}\!=\!1GeV$, $\Theta_{cut}\!=\!7^o$, $\varphi_{cut}\!=\!3^o$).
}\label{mass_llg}
\end{figure}

The mass contribution is small
in the great part of phase space of final particles.
The most significant contribution is for the $J\!=\!0$ energy spectra
(see Fig. \ref{mass_llg}).
The high value of the contribution corresponds
to regions where the differential cross section is minimal.
The mass contribution to the total cross section is below the $1\%$ level
at any realistic set of cuts.
It means that the helicity amplitudes is a good approach
for study the $\gamma\gamma\to l^+l^- \gamma$ process.

\section{Luminosity measurement of $J\!=\!0$ beams.}

For analysis the precision of luminosity measurement \cite{gg_llg_excl} that
can be achieved using the reaction $\gamma\gamma\to f \bar{f} \gamma$,
the most interest are offered by the two kinds of measurement.
The first one is the measuring of beams luminosity with the wide energy spectrum.
The second one is the same measurement for the narrow band around the energy of Higgs boson production.

We use for consideration the following parameters:

1. luminosity
\begin{eqnarray*}
{\cal L} (\sqrt{s'} > 0.8 \sqrt{s'_{\rm max}}) & = & 5.3 \cdot 10^{33} {cm}^{-2}s^{-1},\\
{\cal L} (m_H \pm 1 GeV) & = & 3.8 \cdot 10^{32} {cm}^{-2}s^{-1};
\end{eqnarray*}

2. polarization ${\cal P} \approx 90\%$.

Our calculations allow to
choose the set of cuts with the high $J\!=\!0$ cross section
and high ratio $\sigma_{J\!=\!0}\slash \sigma_{J\!=\!2}$:
$\omega_{cut} \!=\! 20 GeV$,
$E_{f,cut} \!=\! 5 GeV$,
$\Theta_{cut} \!=\! 6^\circ$,
$\varphi_{cut} \!=\! 30^\circ$.
For these cuts the total cross sections have the following values:
\begin{eqnarray*}
\sigma (J=0) & = & 0.82 pb,\\
\sigma (J=2) & = & 1.89 pb.
\end{eqnarray*}

So for the precision of luminosity measurement
in a 2 years run ($2 \cdot 10^7 {\rm s}$) one can obtain:
\begin{eqnarray*}
\frac{\Delta {\cal L}}{{\cal L}}
\left(\sqrt{s'} > 0.8 \sqrt{s'_{\rm max}}\right)
& = & 0.35 \%, \\
\frac{\Delta {\cal L}}{{\cal L}}
\left(m_H \pm 1GeV\right)
& = & 1.3 \%.
\end{eqnarray*}

\section{ Lepton-antilepton production in high energy polarized photons interaction }

The luminosity measurement at $J\!=\!2$ beams will be performed
using the reaction $\gamma\gamma\to l^+l^-$. It has the great
cross section that provides the number of events enough for the
$0.1\%$ precision of luminosity determination.

The main task is to calculate the cross section with maximal
precision. For realization of this purpose we have calculated the
complete one-loop QED radiative corrections to cross section of
$\gamma\gamma\to l^+l^-$ process including the hard photon
bremsstrahlung.

The major feature of $\gamma\gamma\to f \bar{f}$ process is the small value of
cross section if the total angular momentum of $\gamma\gamma-$beams equals zero.

We analyze both the angular spectra and the invariant
distributions of final particles. The angular spectrum of final
leptons is calculated in form $d \sigma \slash d
\cos{\Theta(p_{l},p_{\gamma})}$. It is more convenient to use
Lorentz-invariant results for the experimental reasons. Therefore
we analyze the process $\gamma\gamma\to f\bar{f}[+\gamma]$
including $O(\alpha)$-corrections using the method of covariant calculations \cite{shum,peaking}. The
invariant differential cross section is calculated in the form
\mbox{$d \sigma \slash d (p_{l}-p_{\gamma})^2$} and can be used in the
arbitrary experimental configuration.

The cross section of process \ref{proc3} to be calculated is
\begin{eqnarray}\nonumber
d \sigma = \ddx{2 s} {\left|M_{fi}^{\lambda_1, \lambda_2, e_1{}', e_2{}', [\lambda_k]}\right|}^{2} \cdot d \phi,
\end{eqnarray}
where
\begin{eqnarray}\nonumber
\int \! {A {d \phi_{2[3]}}} =
\ddx{\bra{2 \pi}^{2}}
\cdot \frac{d^{3} p_1'}{2 \E_1'}
\cdot \frac{d^{3} p_2'}{2 \E_2'}
[\cdot \frac{d^{3} k}{\bra{2 \pi}^{3} 2 \omega}]
\cdot \delta \bra{p_1+p_2-p_1'-p_2'[-k]}.
\end{eqnarray}

\begin{figure}
\begin{picture}(300,100)(-30,-20)
\Text(200,5)[c]{+ crossed graphs}
\SetOffset(25, 0)
\Photon(15, 20)(50, 30){1}{5}
\Photon(15, 80)(50, 70){1}{5}
\ArrowLine(85, 20)(50, 30)
\ArrowLine(50, 30)(50, 70)
\ArrowLine(50, 70)(85, 80)
\PhotonArc(50, 50)(10, -90, 90){1}{4.5}

\SetOffset(8, 0)
\Photon(115, 20)(150, 30){1}{5}
\Photon(115, 80)(150, 70){1}{5}
\ArrowLine(185, 20)(150, 30)
\ArrowLine(150, 30)(150, 70)
\ArrowLine(150, 70)(185, 80)
\PhotonArc(150, 70)(20, -90, 17){1}{4.5}

\SetOffset(-8, 0)
\Photon(215, 20)(250, 30){1}{5}
\Photon(215, 80)(250, 70){1}{5}
\ArrowLine(285, 20)(250, 30)
\ArrowLine(250, 30)(250, 70)
\ArrowLine(250, 70)(285, 80)
\PhotonArc(250, 30)(20, -17, 90){1}{4.5}

\SetOffset(-25, 0)
\Photon(315, 20)(340, 30){1}{5}
\Photon(315, 80)(340, 70){1}{5}
\ArrowLine(385, 20)(360, 30)
\ArrowLine(360, 30)(340, 30)
\ArrowLine(340, 30)(340, 70)
\ArrowLine(340, 70)(360, 70)
\ArrowLine(360, 70)(385, 80)
\Photon(360, 30)(360, 70){1}{5}
\end{picture}
\vspace{-20pt}\label{fig_2l_loops}\caption{QED loop corrections.}
\end{figure}
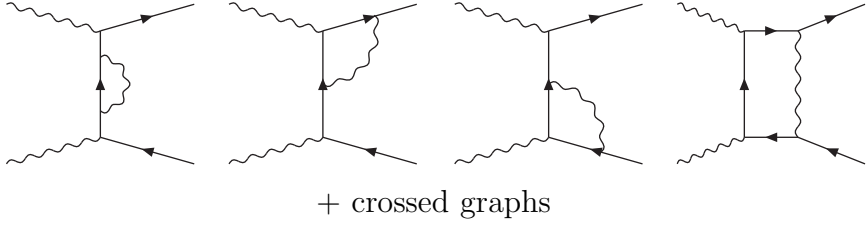
\begin{figure}
\begin{picture}(300,100)(-80,-10)
\SetOffset(10, 0)
\Photon(15, 20)(40, 30){1}{4}
\Photon(15, 80)(40, 70){1}{4}
\ArrowLine(85, 20)(40, 30)
\ArrowLine(40, 30)(40, 70)
\ArrowLine(40, 70)(85, 80)
\Photon(60, 74)(78, 60){1}{4}

\SetOffset(0, 0)
\Photon(115, 20)(140, 30){1}{4}
\Photon(115, 80)(140, 70){1}{4}
\ArrowLine(185, 20)(140, 30)
\ArrowLine(140, 30)(140, 70)
\ArrowLine(140, 70)(185, 80)
\Photon(140, 50)(168, 50){1}{4}

\SetOffset(-10, 0)
\Photon(215, 20)(240, 30){1}{4}
\Photon(215, 80)(240, 70){1}{4}
\ArrowLine(285, 20)(240, 30)
\ArrowLine(240, 30)(240, 70)
\ArrowLine(240, 70)(285, 80)
\Photon(260, 26)(278, 40){1}{4}
\Text(160,5)[c]{+ crossed graphs}

\end{picture}
\vspace{-10pt}\caption{Real photon emission diagrams.}\label{fig_2l_brems}
\end{figure}
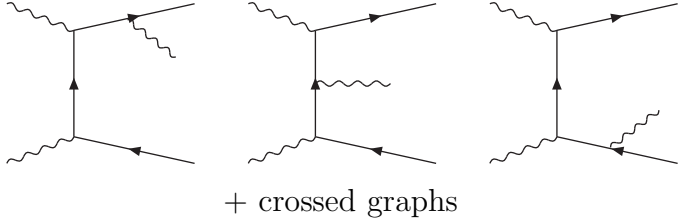

The matrix elements are obtained using the method of helicity ampli\-tudes\mbox{ } \cite{ha}:
\begin{eqnarray}\label{b1}
{\left|M_{2}^{+-+-}\right|}^{2} = 4 e^{4} \frac{u}{t} = 4 e^{4} \frac{1+\cos\Theta_{2,2'}}{1-\cos\Theta_{2,2'}}
= 4 e^{4} \frac{s+t}{-t},
\end{eqnarray}
\begin{eqnarray}\label{b2}
{\left|M_{2}^{+--+}\right|}^{2} = 4 e^{4} \frac{t}{u} = 4 e^{4} \frac{1-\cos\Theta_{2,2'}}{1+\cos\Theta_{2,2'}}
= 4 e^{4} \frac{-t}{s+t}.
\end{eqnarray}

The set of invariants
\begin{eqnarray}\nonumber
\begin{array}{llll}
s = {\left(p_1+p_2\right)}^2,
&
t = {\left(p_2{}'-p_2\right)}^2,
&
u = {\left(p_2{}'-p_1\right)}^2,
&
y = - t \slash s,
\\
\um = 2 \spr{p_1{}'}{k},
&
\vm = 2 \spr{p_2{}'}{k},
&
z = 2 \spr{p_1}{k},
&
z{}' = 2 \spr{p_2}{k}
\end{array}
\end{eqnarray}
are introduced.

It is essential feature of this process that $M_{2}^{++xx}$ and $M_{2}^{--xx}$ amplitudes
at the Born approximation have
the order $\bra{m^2\slash s}^{2}$ and are negligible at high energies.

The integration over $d \phi$ for the $\gamma\gamma\to l^{+} l^{-}$
process is performed as follows:
\begin{eqnarray}
\int {A {d \phi_2}} = \frac{1}{8 \pi s} \int{A \: {d t_{(2)}}}
\simeq \frac{1}{16 \pi} \int{A \: {d \cos{\Theta_{2,2'}}}}.
\end{eqnarray}

The QED loop corrections are represented by diagrams on Fig. \ref{fig_2l_loops}.
We can factorize them upon the Born cross section as follows:

\begin{eqnarray}
d \sigma^{+-xx} = d \sigma_{(2)}^{+-xx}\cdot \frac{\alpha}{2 \pi}\cdot \delta_{V}^{+-xx},
\end{eqnarray}
\begin{eqnarray}\label{v2}
\begin{array}{c}
{\displaystyle
\delta_{V}^{+-+-} =
2 \ln{\frac{s}{\lambda^{2}}} \bra{1-\ln{\frac{s}{\m2}}}
+\ln^{2}{\frac{s}{\m2}}+\ln{\frac{s}{\m2}}+\ln^{2}{\frac{-u}{s}}+
}
\\ \\
{\displaystyle 
+\frac{s^2}{u^2} \ln^{2}{\frac{-t}{s}}
+\bra{1-\frac{2 s}{u}}\ln{\frac{-t}{s}}
+\frac{4 \pi^{2}}{3}-3;
}
\end{array}
\end{eqnarray}
\begin{eqnarray}\label{v3}
\begin{array}{c}
{\displaystyle
\delta_{V}^{+--+} =
2 \ln{\frac{s}{\lambda^{2}}} \bra{1-\ln{\frac{s}{\m2}}}
+\ln^{2}{\frac{s}{\m2}}+\ln{\frac{s}{\m2}}+\ln^{2}{\frac{-t}{s}}+
}
\\ \\
{\displaystyle 
+\frac{s^2}{t^2} \ln^{2}{\frac{-u}{s}}
+\bra{1-\frac{2 s}{t}}\ln{\frac{-u}{s}}
+\frac{4 \pi^{2}}{3}-3.
}
\end{array}
\end{eqnarray}

Here we have introduced the finite photon mass $\lambda$
to remove the infrared divergence.

The real photon emission for this process is a pure QED reaction.
It is indistinguishable from the
$\gamma\gamma\to l^{+} l^{-}$ process in the infrared (IR) limit
and it's singularities cancel
ones caused by loop corrections.

The integration over $d \phi$ leads to (in non-covariant expressions the c.m.s. system is used) \cite{shum,peaking}
\begin{eqnarray}\label{br2}
\begin{array}{l}
{\displaystyle
\int{A {d \phi_3}}
= \ddx{2 {4 \pi}^{3}} \int{\Jint{A} \: {d \um} \: {d y}}
= \ddx{4 {4 \pi}^{3} s} \int{\Jint{A} \psi_{\um} \: {d \um} \: {d \cos{\Theta_{2,2'}}}},
}
\end{array}
\end{eqnarray}
here
\begin{eqnarray}\label{br2a}
\Jint{A}=\frac{1}{\pi} \int{\frac{d^3 k}{\omega} \: A \: \delta\left(Q^2-\m2-2 Q_{0}\omega\right) \Theta\left(Q_{0}-\omega\right)},
\end{eqnarray}
\begin{eqnarray}\nonumber
Q=p_1+p_2-p_2{}'.
\end{eqnarray}

Using the method of helicity amplitudes \cite{ha}, one can calculate
\begin{eqnarray}\label{br3}
{\left|M^{+--++}\right|}^{2} = e^6
\frac{4 \spr{p_1'}{p_2'} \bra{\spr{p_2'}{p_2}}^{2}}{\spr{p_1'}{k} \spr{p_2'}{k} \spr{p_1'}{p_1} \spr{p_2'}{p_1}},
\end{eqnarray}

The other non-vanishing amplitudes are
obtained from $\left|M^{+--++}\right|$
by using C, P, Bose and crossing (between final and initial particles) symmetries:
$$
d \sigma^{-\lambda_1,-\lambda_2,-e_1{}',-e_2{}',-\lambda_3} =
d \sigma^{\lambda_1,\lambda_2,e_1{}',e_2{}',\lambda_3},
\;\, (P) \qquad\quad\;
$$
$$
\qquad\qquad\:\:\,
d \sigma^{+-+--} =
d \sigma^{+--++} {}_{\mid_{1\leftrightarrow 2}}, \;\:\, (P+Bose)
$$
$$
\quad\,
d \sigma^{+-+-+} =
d \sigma^{+--++} {}_{\mid_{1'\leftrightarrow 2'}}, \;(C)
$$ $$
\qquad\qquad\quad\:\:
d \sigma^{+--+-} =
d \sigma^{+--++} {}_{\mid_{1\leftrightarrow 2 \atop
1'\leftrightarrow 2'}}, \!\;(CP+Bose)
$$
$$
\qquad\qquad\qquad\:
d \sigma^{+++--} =
d \sigma^{+--++} {}_{\mid_{3\leftrightarrow 2 \atop
1'\leftrightarrow 2'}}, \!\;(C+crossing)
$$
$$
\quad\;\!
d \sigma^{++-+-} =
d \sigma^{+++--} {}_{\mid_{1'\leftrightarrow 2'}}.
\;\;\! (C)
$$

The last couple of substitutions leads to the non-divergent leading term of $\gamma^+ - \gamma^+$ -- scattering.

It is convenient to perform the integration over the phase-space
of the final particles numerically. But the Monte-Carlo methods of
numerical analysis \cite{mc} require to eliminate all the divergences in the
integration expressions.

The "forward-backward" divergences can be deleted
by imposing cuts on the scattering angle (in calculation of $d \sigma \slash d \cos{\Theta}$)
or on the $(\spr{p_i}{p_f})$-invariants (for $d \sigma \slash d y$).
Another singularities should be extracted as a single expression $ {\left|M\right|}^{2}_{sub}$.
After this term has been subtracted the matrix element doesn't contain divergences
and can be integrated numerically.
The singular term $ {\left|M\right|}^{2}_{sub}$ should be integrated analytically.

The infrared behaviour of helicity amplitudes
can be found by covariant expanding (\ref{br3}) of matrix elements into
a series around pole $\omega_{\gamma_{real}} \to 0$:

\begin{eqnarray}\label{br5}
\begin{array}{l}
{\displaystyle
{\left|M^{+-+-}\right|}^{2}_{IR} = 16 e^6
\frac{s}{\um \vm} \frac{u}{t}
\bra{1-\frac{\vm}{s}-\frac{\um}{s}}
+ 8 e^6 \frac{s}{\um \vm}
\frac{(\um-z) u + (\vm-z) t}{t^2},
}
\\
{\displaystyle
{\left|M^{+--+}\right|}^{2}_{IR} = 16 e^6
\frac{s}{\um \vm} \frac{t}{u}
\bra{1-\frac{\vm}{s}-\frac{\um}{s}}
- 8 e^6 \frac{s}{\um \vm}
\frac{(\um-z) u + (\vm-z) t}{u^2}.
}
\end{array}
\end{eqnarray}

The first term of each expression
has the usual IR-singularity
and the rest one is divergent in the massless limit.

The divergences caused by $\spr{p_f}{k} \to 0$
can be extracted \cite{peaking} using the method of peaking approximation:
\begin{eqnarray}\label{br6}
\begin{array}{l}
{\displaystyle
{\left|M^{+-+-}\right|}^{2}_{peak} = 8 e^6
\frac{s}{\um\vm} \frac{u}{t}
\bra{1-\frac{\vm}{s}+\frac{\vm^2}{s^2}-\frac{\um}{s}+\frac{\um^2}{s^2}},
}
\\
{\displaystyle
{\left|M^{+--+}\right|}^{2}_{peak} = 8 e^6
\frac{s}{\um\vm} \frac{t}{u}
\bra{1-\frac{\vm}{s}+\frac{\vm^2}{s^2}-\frac{\um}{s}+\frac{\um^2}{s^2}}.
}
\end{array}
\end{eqnarray}

Each formula of eqs. (\ref{br5}) and (\ref{br6}) can be combined into the united expression:
\begin{eqnarray}\label{br7}
\begin{array}{l}
{\displaystyle
{\left|M^{+-+-}\right|}^{2}_{sub} =
8 e^6 \frac{s}{\um\vm}
\bra{
\frac{u}{t}
\bra{1-\frac{\vm}{s}+\frac{\vm^2}{s^2}-\frac{\um}{s}+\frac{\um^2}{s^2}}
+ \frac{t \u1-u \t1}{t^{2}} },
}
\\
{\displaystyle
{\left|M^{+--+}\right|}^{2}_{sub} = 8 e^6 \frac{s}{\um\vm}
\bra{
\frac{t}{u}
\bra{1-\frac{\vm}{s}+\frac{\vm^2}{s^2}-\frac{\um}{s}+\frac{\um^2}{s^2}}
- \frac{t \u1-u \t1}{u^{2}} }.
}
\end{array}
\end{eqnarray}

This directly leads to eqs. (\ref{br5}) in the IR-limit.
And it differs from (\ref{br6}) on the term that vanishes in the peaking limits due to
\begin{eqnarray}\label{br7b}
(\um-z)u+(\vm-z)t =
\bra{t \u1-u \t1}
\, \stackrel{peak}{\longrightarrow} \, 0.
\end{eqnarray}

The analytical integration of (\ref{br7}) over the phase-space is
performed according to (\ref{br2}). The second term in (\ref{br7})
is only a IR-divergent one.
To simplify further calculations we introduce
arbitrary value $\bar{\um}$ as an upper bound for it's integration
(and subtraction). Neither numerical no analytical part of the
result does not depend on $\bar{\um}$ if it is chosen in the
region \mbox{$\m2 \! \ll \! \bar{\um} \! \ll \! s$} (or \mbox{$\m2 \! \ll \! \bar{\um} \! \ll \! \bra{s+t}$}
in case of $y$-dependent differential cross-section).

The IR-divergences can be factorized upon matrix element in a covariant path as follows
\begin{eqnarray}\label{br_l1}
M^{\lambda} = e M^{Born} \bra{\frac{{p_1{}'}_{\mu}}{\spr{p_1{}'}{k}}-\frac{{p_2{}'}_{\mu}}{\spr{p_2{}'}{k}}}
\epsilon_{k}^{\mu},
\end{eqnarray}
that after squaring gives
\begin{eqnarray}\label{br_l2}
{|M^{\lambda}|}^{2} =
4 e^{2}
{|M^{Born}|}^{2}
\bra{\frac{s{}'}{\um \vm}-\frac{m^2}{\um^2}-\frac{m^2}{\vm^2}}.
\end{eqnarray}
The $m^2$-dependent terms form (\ref{br_l2}) do not appear in helicity amplitude expressions
since setting mass to zero
but they should be included in calculations for proper cancelation of divergences.

The result of analitical integration over the phase space of final photon
for the "soft" and "collinear" parts of bremsstrahlung is
\begin{eqnarray}\label{br12}
\delta_{R}^{\Theta} =
2 \ln{\frac{s}{\lambda^{2}}} \bra{\ln{\frac{s}{\m2}}-1}
-\ln^{2}{\frac{s}{\m2}}
-\ln{\frac{s}{\m2}}
-\frac{4 \pi^{2}}{3}+\frac{13}{2}.
\end{eqnarray}

Combining loop correction expressions (\ref{v2}, \ref{v3})
and the bremsstrahlung contribution (\ref{br12}) one can obtain
\begin{eqnarray}\label{br13}
\delta_{R}^{\Theta} + \delta_{V}^{+-+-} =
\ln^{2}\bra{1-y}+\ddx{(1-y)^{2}} \ln^{2}{y}+\bra{1+\frac{2}{1-y}}\ln{y}+\frac{7}{2},
\end{eqnarray}
\begin{eqnarray}\label{br14}
\delta_{R}^{\Theta} + \delta_{V}^{+--+} =
\ln^{2}{\bra{y}}+\ddx{y^{2}} \ln^{2}\bra{1-y}+\bra{1+\frac{2}{y}}\ln{\bra{1-y}}+\frac{7}{2}.
\end{eqnarray}

Here $y$ is the function of angle between initial and final particles:
\begin{eqnarray}\nonumber
y = \ddx{2}\bra{1-\cos{\Theta_{2,2{}'}}}.
\end{eqnarray}

The integration results
for the invariant-dependent spectra
are so complicated that
can't be outlined here.

The final state polarization can scarcely be measured at experiment.
That is the reason for summarizing over the helicities of all final particles.

\begin{figure}[!ht]
\leavevmode
\begin{minipage}[h]{.5\linewidth}
\centering
\includegraphics[width=\linewidth, height=2.4in, angle=0]{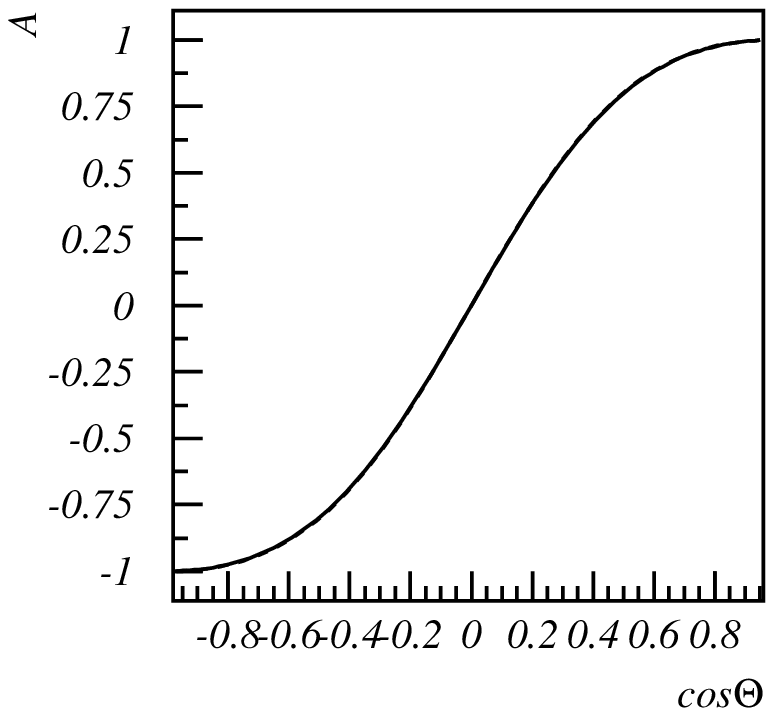}
\end{minipage}\hfill
\begin{minipage}[h]{.5\linewidth}
\centering
\includegraphics[width=\linewidth, height=2.4in, angle=0]{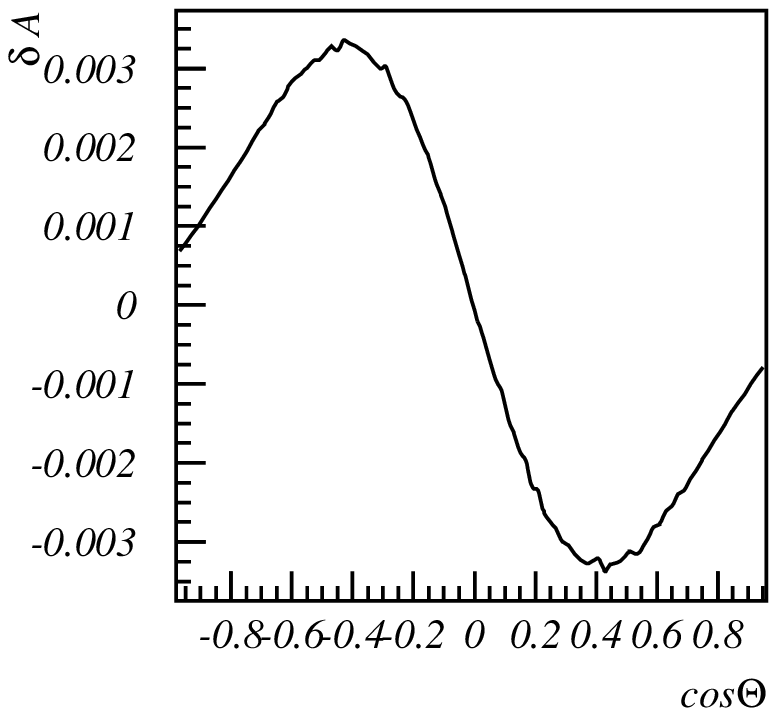}
\end{minipage}
\vspace{-20pt}\caption{
Angular-dependent polarization asymmetry and QED correction contribution
$\delta (A) = A_{tot} -  A_{Born}$.
}\label{ang_asym}
\end{figure}

\begin{figure}[!ht]
\leavevmode
\begin{minipage}[h]{.5\linewidth}
\centering
\includegraphics[width=\linewidth, height=2.4in, angle=0]{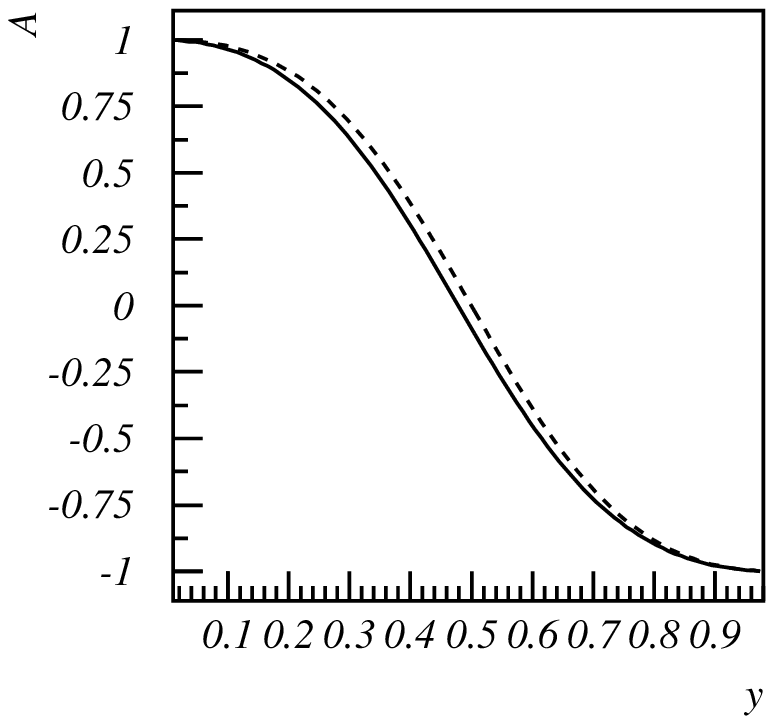}
\end{minipage}\hfill
\begin{minipage}[h]{.5\linewidth}
\centering
\includegraphics[width=\linewidth, height=2.4in, angle=0]{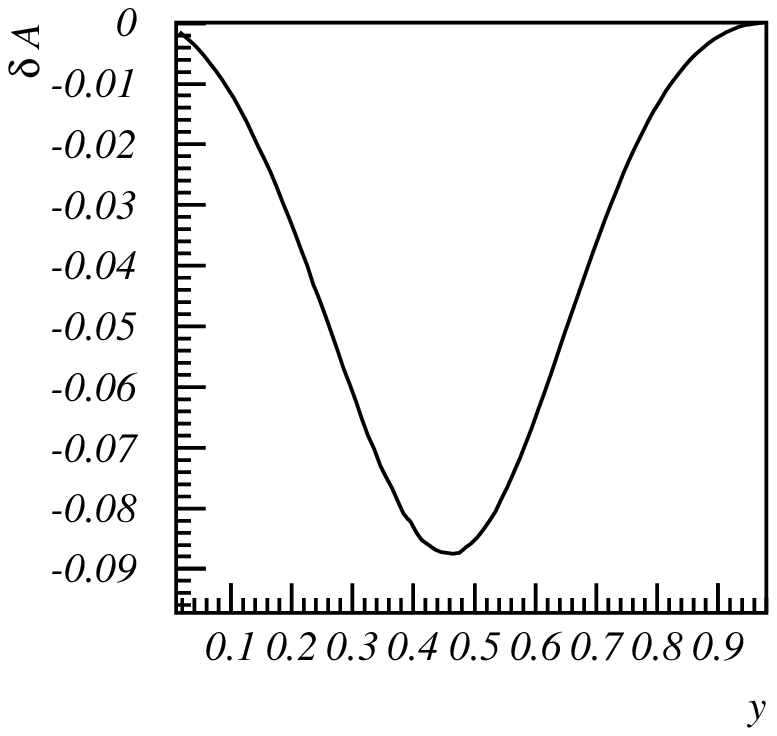}
\end{minipage}
\vspace{-20pt}\caption{
Polarization asymmetry and QED correction contribution
$\delta (A) = A_{tot} -  A_{Born}$.
}\label{inv_asym}
\end{figure}

We present here plots for polarization asymmetries and ${\cal O} (\alpha)$-correction to it
(see Figs. \ref{ang_asym}, \ref{inv_asym}).
The graphs are composed for c.m.s. energy \mbox{$\sqrt{s} \! = \! 120 \; GeV$}
(the energy of supposed resonant Higgs boson pro\-duc\-tion \cite{gg_h}).

The major feature of $\gamma\gamma\to l^{+} l^{-}$ process is the small value of
cross section if the total angular momentum of $\gamma\gamma$-beams equals zero.
This polarization selectivity can be useful at the experiment.


For the measurement the luminosity of $J\!=\!2$ beams
one will use the events of $\gamma\gamma\to l^+l^-$ process.
The precision of measurement the luminosity of $J\!=\!2$ beams
that can be achieved using $\gamma\gamma\to l^+l^-$ process
can be calculated in the same way that one for $J\!=\!0$ beams.
We introduce the $\omega_{max}$ parameter for
the maximal energy of bremsstrahlung photon
that will still result
the detection of single exclusive $\gamma\gamma\to l^+l^-$ event.
For the supposed detector parameters
($\omega_{max} \!=\! 1 GeV$,
$E_{f,cut} \!=\! 1 GeV$,
$\Theta_{cut} \!=\! 7^\circ$) one can obtain:

\begin{eqnarray*}
\frac{\Delta {\cal L}}{{\cal L}}
\left(\sqrt{s'} > 0.8 \sqrt{s'_{\rm max}}\right)
& = & 0.04 \%, \\
\frac{\Delta {\cal L}}{{\cal L}}
\left(m_H \pm 1GeV\right)
& = & 0.1 \%.
\end{eqnarray*}

The achieved precision is sufficient for the huge variety of
experiments at the photon collider.


\section{Boson production in $\gamma\gamma$-collisions}

    Future high-energy linear $e^+ e^-$ colliders in $\gamma e$ and
$\gamma\gamma$ mode could be a  very useful instrument to explore
mechanism of symmetry breaking in electroweak interaction using self
couplings test of the $W$ and $Z$ bosons in non-minimal gauge
models. $WW$-production would be provided mainly by $\gamma \gamma
$-scattering  \cite{r7}. The Born cross section
$\sigma(\gamma\gamma\rightarrow W^+ W^-)$ is about 110pb at 1 TeV on
unpolarized $\gamma$-beams. Corresponding cross section of
$WW$-production on electron colliders is an order of magnitude
smaller and amounts to 10pb. One needs to consider a reaction
$\gamma\gamma\rightarrow W^+W^-Z$ since its cross section becomes
about 5\%-10\% of the cross section $WW$-production at energies
$\sqrt{s}\geq 500$ GeV. The anomalous three-linear  \cite{r8} $\gamma
W W$ and $Z W W$ and quartic  \cite{r9} $\gamma\gamma W W$, $\gamma Z
W W$, $Z Z W W$ etc. couplings induce deviations of the lowest-order
cross section from the Standard Model.

In order to evaluate contributions of anomalous couplings a cross
section of $\gamma\gamma\rightarrow W^+ W^-$ must be calculated with
a high precision and extracted from experimental data. Therefore one
needs to calculate the main contribution of high order electroweak
effects: one-loop correction, real photon and $Z$ emission.

Lagrangian of three-boson $(WW\gamma$ and $WWZ)$ interaction in the
most general form can be presented as
\begin{gather}
 L_{WWV}=-g_{WWV} i [g_1^{V}\left(W_{\mu\nu}^+ W^{\mu} V^{\nu}-W_{\mu}^+ V_{\nu}
W^{\mu \nu}\right)\phantom{]}\nonumber\\
+k_{V} W_{\mu}^+ W^{\nu} V^{\mu \nu}+ i \lambda_V / m_W^2
W_{\lambda\mu}^+ W^{\mu} V^{\nu\lambda}-\nonumber\\
-g_4^V W_{\mu}^+ W^{\nu}\left(\partial^{\mu} V^{\nu}+\partial^{\nu}
V^{\mu}\right)+\\
\phantom{[} +g_5^V \epsilon^{\mu\nu\rho\sigma}\left(W_{\mu}^+
\overrightarrow{\partial}_{\rho}\!\!\!\!\!\!\!\!\!\overleftarrow{\phantom{\partial}}\;\; W_{\nu}\right) V_{\sigma}+\nonumber\\
+i k_V W_{\mu}^+ W_{\nu} \tilde{V}^{\mu \nu}+i \tilde{\lambda}_V
/m_W^2 W_{\lambda\mu}^+ W^{\mu}_{\nu} V^{\nu\lambda}].\nonumber
\end{gather}
Here $V_{\mu}$ is the photon or $Z$-boson field (correspondingly,
$V=\gamma$ or $V=Z$), $W_{\mu}$ -- $W^{-}$-boson field,
\begin{equation}
W_{\mu\nu}=\partial_{\mu} W_{\nu}- \partial_{\nu}
W_{\mu},\;\;V_{\mu\nu}=\partial_{\mu} V_{\nu}- \partial_{\nu}
V_{\mu},
\end{equation}
$\tilde{V}_{\mu\nu}=\frac{1}{2}\epsilon_{\mu\nu\rho\sigma}
V^{\rho\sigma}$ and $A\overrightarrow{\partial}_{\mu}
\!\!\!\!\!\!\!\!\!\overleftarrow{\phantom{\partial}}\;\;B=A(\partial_{\mu}B)-(\partial_{\mu}A)B$.
The parameter of interaction $g_{WWV}$ are fixed as follows:
\begin{equation}
 g_{WW\gamma}=e,\; g_{WWZ}=e \cos{ \theta_W}.
\end{equation}

In case of $WW\gamma$-interaction the first term corresponds to the
minimal interaction (in case of $g_1^\gamma = 1$). The parameters of
the second and third terms are connected with magnetic momentum and
quadrupole electric one of $W$-boson correspondingly as
\begin{equation}\mu_W=\frac{e}{2 m_W}(1+k_\gamma
+\lambda_\gamma),\;Q_W=\frac{e}{m_W^2}(\lambda_\gamma - k_\gamma).
\end{equation}
The last two operators parameters are connected with electric dipole
moment $d_W$ as well as quadrupole magnetic moment $\tilde{Q}_W$:
\begin{equation}
    d_W=\frac{e}{2 m_W}(\tilde{k}_\gamma +\tilde{\lambda}_\gamma),\;\tilde{Q}_W=\frac{e}{m_W^2}(\tilde{\lambda}_\gamma -
    \tilde{k}_\gamma).
\end{equation}

In frame of the SM $W W \gamma$- and $W W Z$-vertices are determined
by gauge group $SU(2)\otimes U(1)$. In the lowest order of
perturbative theory only $C$- and $T$-invariant corrections exist
(in this case $k_V=1$, $\lambda_V=0$). However electroweak radiative
corrections (loop diagrams with heavy charged fermions) can give
significant contribution in $k_V$ and $\lambda_V$ as well as $C$-
and $T$-violate interaction.

There are four-boson vertices giving additional independent
information about gauge structure of electroweak interaction. The
corresponding cross sections give contribution in cross section of
boson production in $e \gamma$- and $\gamma\gamma$-scattering.

If we will consider only the interactions which conserve  $P$- and
$C$-symmetry, Lagrangian four-boson interaction includes two
6-dimension operators
\begin{equation}\label{lq6}
\begin{gathered}
L_{Q}^{(6)}=-\frac{\pi\alpha}{4m_W^2}\left[a_{o}F_{\alpha\beta}F^{\alpha\beta}\left(\vec{W}_{\mu}\cdot\vec{W}^{\mu}
\right)
+\right.\\\left.+a_{c}F_{\alpha\mu}F^{\alpha\nu}\left(\vec{W}^{\mu}\cdot\vec{W}_{\nu}
\right)\right],
\end{gathered}
\end{equation}
where $F_{\alpha\beta}$ -- tensor of electromagnetic field,
$\vec{W}_\mu$ represent $W$-triplet, $a_0$ and $a_c$ -- anomalous
constants. The first term  corresponds to neutral scalar exchange.
One-loop corrections due to charged heavy fermions give
contributions with four-boson vertices to the both terms of the
Lagrangian (\ref{lq6}).

Charged scalars give contribution proportional to $a_0$ only.

Since cross section of photoboson production rises to constant value
and cross section of  electron-positron interaction decreases with
energy growth as reverse proportional dependence $s^{-1}$ when
central mass is equal to 500 GeV, the photoproduction of boson cross
section is an order bigger than  $e^+ e^-$ interaction cross section
and is the most important source of information about anomalous
boson couplings.

We have considered the anomalous quartic boson vertices. For this
purpose the following 6-dimensional $SU(2)_C$ Lagrangian
 \cite{r9,r10} have been chosen:
\begin{equation}
\begin{gathered}
L_0=-\frac{e^2}{16\Lambda^2}a_0 F^{\mu\nu} F_{\mu\nu}
\vec{W}^{\alpha}\cdot\vec{W}_{\alpha},\\
L_c=-\frac{e^2}{16\Lambda^2}a_c F^{\mu\alpha} F_{\mu\beta}
\vec{W}^{\beta}\cdot\vec{W}_{\alpha},\\
\tilde{L}_0=-\frac{e^2}{16\Lambda^2}\tilde{a}_0 F^{\mu\alpha}
\tilde{F}_{\mu\beta} \vec{W}^{\beta}\cdot\vec{W}_{\alpha}.
\end{gathered}
\end{equation}
where the triplet gauge boson $\vec{W}_\mu$ and the field-strength
tensors
\begin{equation*}
\begin{gathered}
F_{\mu \nu} =\partial_{\mu}A_\nu - \partial_\nu
A_\mu,\;W_{\mu\nu}^i=\partial_\mu W_\nu^i -\partial_\nu W_\mu^i,\\
\tilde{F}_{\mu\nu}=\frac{1}{2}\varepsilon_{\mu \nu\rho\sigma}F^{\rho
\sigma}
\end{gathered}
\end{equation*}
are introduced. As one can see the operators $L_0$ and $L_c$ are
$C$-, $P$- and $CP$-invariant. $\tilde{L}_0$ is the $P$- and
$CP$-violating operator.

\begin{figure}[!h]
     \leavevmode
\centering
\includegraphics[width=2.4in]{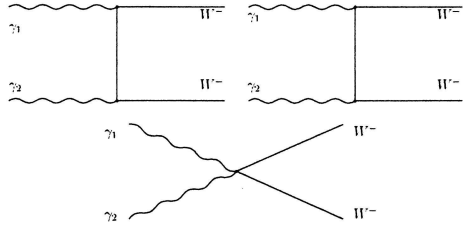}
\vspace{-10pt}\caption{The Feynman diagrams for $W^+W^-$-production}
\end{figure}

We start from the explicit expression for the amplitude of the
process $\gamma\gamma\rightarrow W^+ W^-$
\begin{equation}
    M=G_\nu \varepsilon_\mu (k_1) \varepsilon_\nu (k_2) \varepsilon_\alpha
    (p_+) \varepsilon_\beta (p_-) M_T^{\mu\nu\alpha\beta},
\end{equation}
where
\begin{equation}
M_T^{\mu\nu\alpha\beta}=\sum_{i=1}^{3}M_i^{\mu\nu\alpha\beta},\label{Mt}
\end{equation}
$k_1$, $k_2$, $p_+$, $p_-$ are four-momenta for the $\gamma$,
$\gamma$, $W^+$, $W^-$ and $\varepsilon_\mu (k_1)$, $\varepsilon_\nu
(k_2)$, $\varepsilon_\alpha (p_+)$, $ \varepsilon_\beta (p_-)$ their
polarizations  respectively,
\begin{equation*}
    G_\nu=e^3 \cot \theta_W .
\end{equation*}

\begin{figure}[!h]
     \leavevmode
\centering
\includegraphics[width=2.2in,height=2.2in]{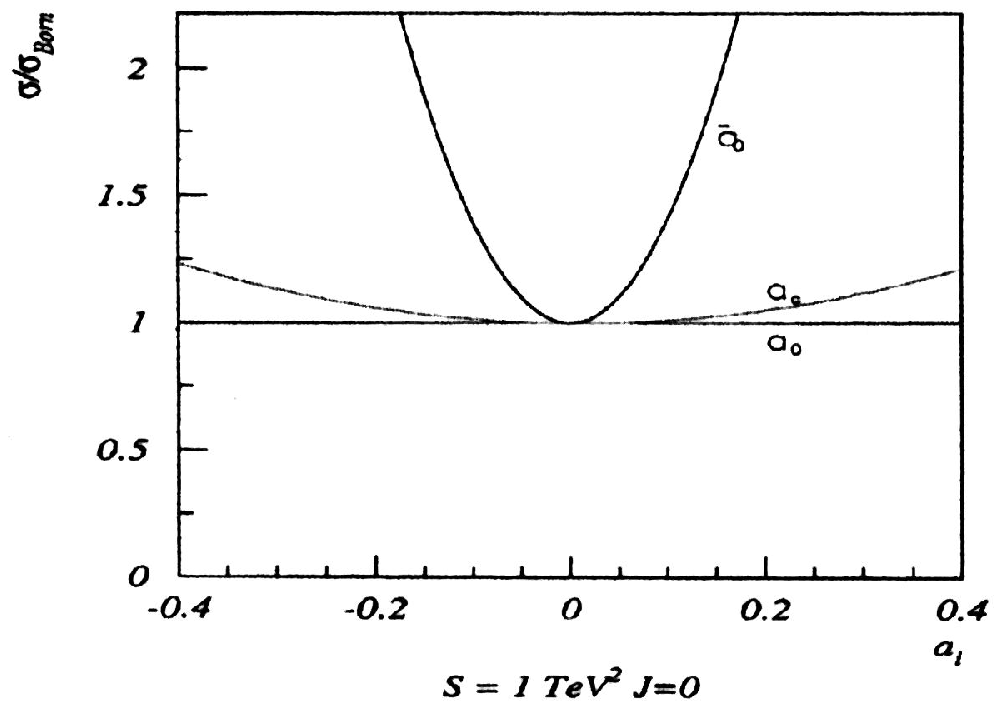}
\vspace{-10pt}\caption{Dependencies of the ratio $\sigma/\sigma_{SM}$ on the
various couplings}     \label{fig7}
\end{figure}

\begin{figure}[!h]
     \leavevmode
\centering
\includegraphics[width=2.2in,height=2.2in]{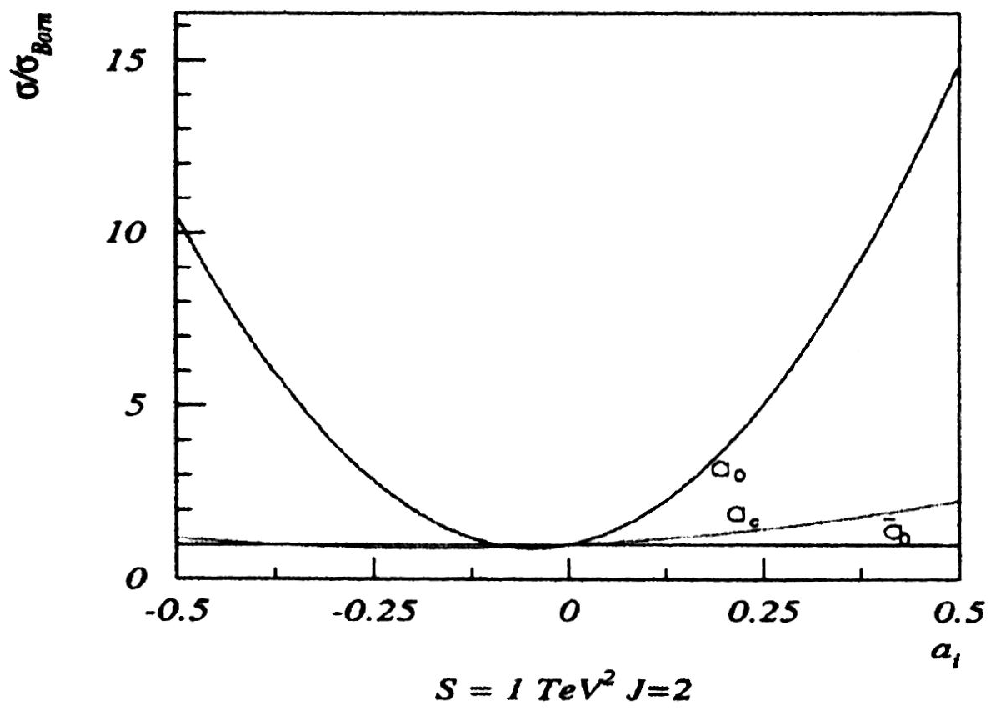}
\vspace{-10pt}\caption{Dependencies of the ratio $\sigma/\sigma_{SM}$ on the
various couplings}     \label{fig8}
\end{figure}

\begin{figure}[!h]
     \leavevmode
\centering
\includegraphics[width=2.2in,height=2.2in]{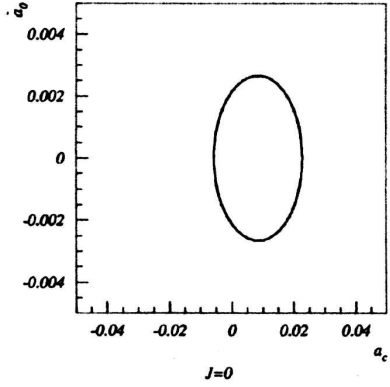}
\vspace{-10pt}\caption{Contour plots on $(a_c,\tilde{a}_0)$ for $1\delta$ at $J=0$
photon beams}\label{fig9}
\end{figure}

\begin{figure}[!h]
     \leavevmode
\centering
\includegraphics[width=2.2in,height=2.2in]{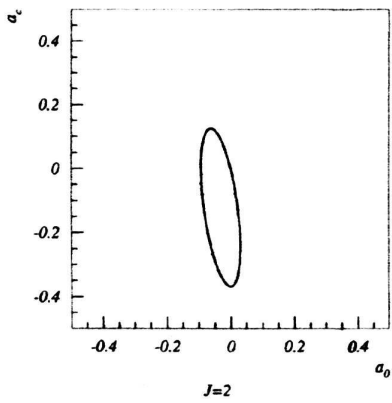}
\vspace{-10pt}\caption{Contour plots on $(a_0,a_c)$ for $1\delta$ at $J=2$ photon
beams}\label{fig10}
\end{figure}

\begin{figure}[!h]
     \leavevmode
\centering
\includegraphics[width=2.2in,height=2.2in]{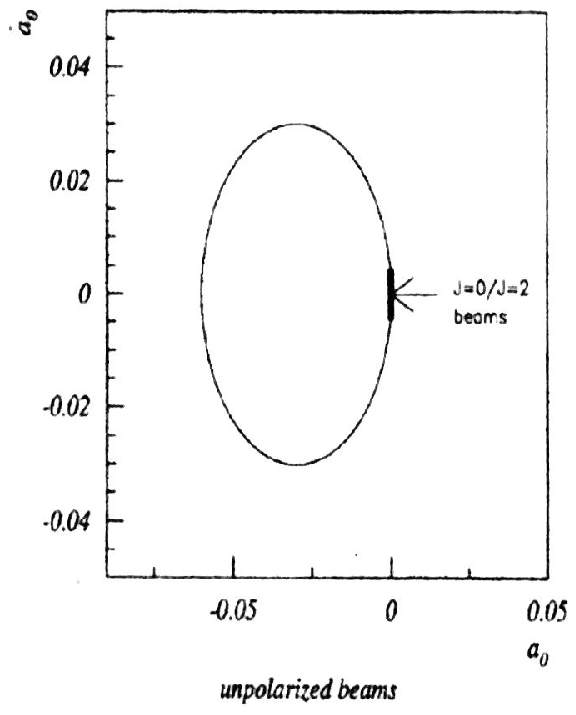}
\vspace{-10pt}\caption{Contour plots on $(a_0,\tilde{a}_0)$ for $1\delta$}\label{fig11}
\end{figure}

\begin{figure}[!h]
     \leavevmode
\centering
\includegraphics[width=2.2in,height=2.2in]{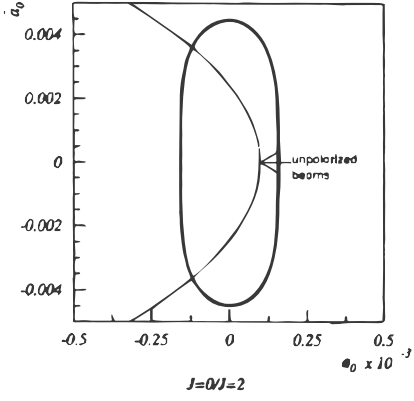}
\vspace{-10pt}\caption{Contour plots on $(a_0,\tilde{a}_0)$ for $1\delta$}\label{fig12}
\end{figure}

Total cross section of $\gamma\gamma$-boson production can be
presented as
\begin{equation}
    \sigma=\frac{1}{2s}\sum_{\lambda_1\lambda_2\lambda_3\lambda_4}{\int{|M_{\lambda_1\lambda_2\lambda_3\lambda_4}|^2
    d\Gamma}},
\end{equation}
where $M_{\lambda_1\lambda_2\lambda_3\lambda_4}$ have been defined
by eq. (\ref{Mt}), $d\Gamma$ is phase space element of the bosons. The
dependence of total cross section $\sigma(W^+W^-)$ on anomalous
parameters was investigated at the following  experimental
conditions:\\
--~The center-of-mass energy of $\gamma\gamma(\sqrt{s})$ in
$\gamma\gamma\rightarrow W^+ W^-$ is fixed at 1 TeV;\\
--~Photon luminosity $L$ is supposed to be 100
$\text{fb}^{-1}$/year;\\
--~In ILC experiments for $\gamma\gamma$-scattering polarization
states of the photon beams will be fixed by $J=0$ or $J=2$ states;
--~In addition it is assumed that the final $W$-bosons will be
detected with certain polarization states; \\
and the results are presented in Figs. \ref{fig7}--\ref{fig12}.


\begin{figure}[!h]
     \leavevmode
\centering
\includegraphics[width=2.4in]{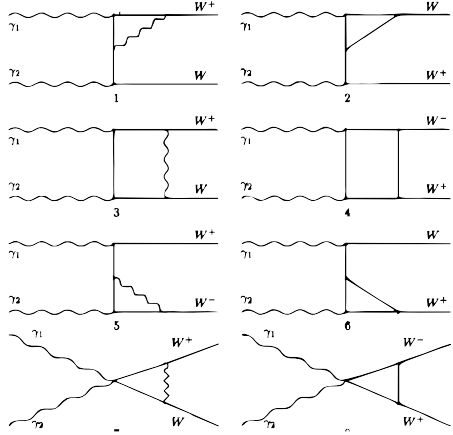}
\vspace{-10pt}\caption{The Feynman diagrams of one-loop amplitudes of the
$\gamma\gamma \rightarrow W^+W^-$}     \label{oneloop1}
\end{figure}

\begin{figure}[!h]
      \leavevmode
\centering
\includegraphics[width=2.4in]{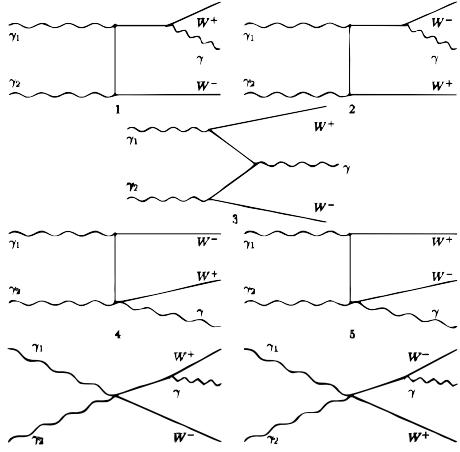}
\vspace{-10pt}\caption{The Feynman diagrams of $\gamma\gamma \rightarrow W^+W^-$
accompanied real photon emission}      \label{oneloop2}
\end{figure}

It is evident that minima of the curves are close to the Standard
Model point $a_i=0$ since the interference between anomalous and
standard part of cross section is very small. Through the region of
$a_i$ is small (about 0.05) the cross section with anomalous
constants may reach values of $1.6\sigma$. Taking into account a
luminosity of photons and beams energy statistical error will be
equal to 0.05 \%. Therefore for precision analysis of experimental
data it is important to calculate radiative corrections. We
calculate $\mathcal{O}(\alpha)$ radiative correction giving maximal
contribution to cross section value. It includes real photon
emission as well as a set of one-loop diagrams (see. Fig. \ref{oneloop1}--\ref{oneloop2}).
Since of ILC-beams energy exceeds the threshold of three boson
production this process must be considered as radiative effect too:
\begin{equation*}
\begin{gathered}
d\sigma(\gamma\gamma \rightarrow W^+ W^-)=\\d\sigma^{\text{Born}}
    (\gamma\gamma \rightarrow W^+
    W^-)+
\end{gathered}
\end{equation*}
\begin{equation}
\begin{gathered}
+\frac{1}{s}\Re(M^\text{Born}M^{1-\text{loop*}})d\Gamma^{(2)} +\\+ d
\sigma^{\text{soft}}(\gamma\gamma \rightarrow W^+
    W^-\gamma)+\\
    + d \sigma^{\text{hard}}(\gamma\gamma \rightarrow W^+
    W^-\gamma)+\\ +d \sigma^{Z}(\gamma\gamma \rightarrow W^+
    W^-Z).
\end{gathered}
\end{equation}
Here $d \sigma^{\text{soft}}(\gamma\gamma \rightarrow W^+
W^-\gamma)=d \sigma^{\text{Born}}(\gamma\gamma \rightarrow W^+
W^-)R^{\text{soft}}(\omega)$, where $\omega$ is soft photon energy
cutoff,
\begin{equation}
\begin{gathered}
    R^\text{soft}=\frac{2\alpha}{\pi}\left\{
\left[-1+\frac{1}{\beta}\left(1-\frac{2m_W^2}{s}\right)\ln{\frac{1+\beta}{1-\beta}}\right]\right.\times\\\times\left.
\left[\ln{2\omega}+\frac{1}{n-4}-\ln(2\sqrt{\pi}+\frac{C}{2})\right]+\right.\\\left.+\frac{1}{2\beta}\ln{\frac{1+\beta}{1-\beta}}+
\frac{1}{2\beta}\left(1-\frac{2m_W^2}{s}\right)\left(\mathop{Spence}{\frac{-2\beta}{1-\beta}}-\right.\right.\\\left.\left.-\mathop{Spence}{\frac{2\beta}{1-\beta}}\right)\right\},\\
\beta=\sqrt{1-4m_W^2/s}.
\end{gathered}
\end{equation}
The differential cross section of hard photon emission is given by
\begin{equation}
\begin{gathered}
    d \sigma^{\text{hard}}(\gamma\gamma \rightarrow W^+
    W^-\gamma)=\\d \sigma(\gamma\gamma \rightarrow W^+
    W^-\gamma)-\\-d \sigma^{\text{soft}}(\gamma\gamma \rightarrow W^+
    W^-\gamma)
\end{gathered}
\end{equation}
and can not be factorized. $d \sigma^{\text{soft}}$ and $d
\sigma^{\text{hard}}$ are independent from infrared divergence and
from cutoff parameter.

\begin{figure}[!h]
     \leavevmode
\centering
\includegraphics[width=2.2in]{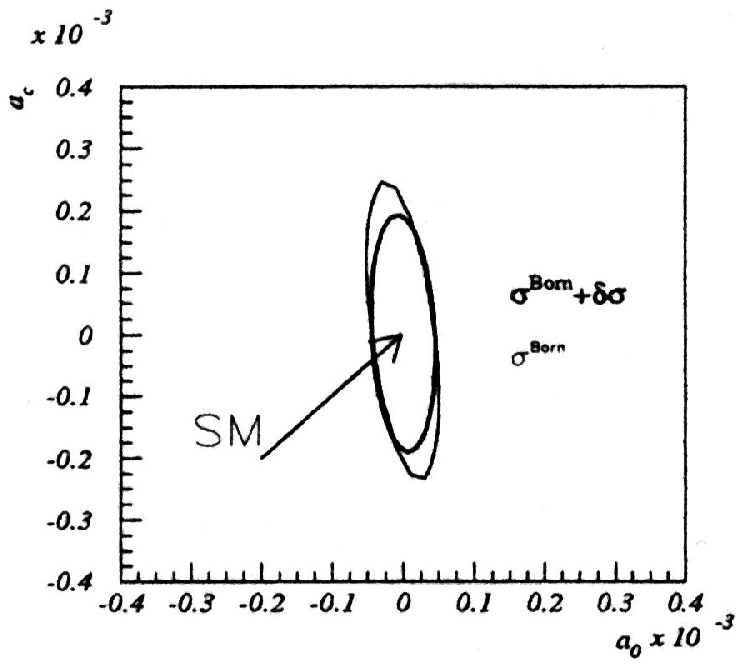}
\vspace{-10pt}\caption{Contour plots on $(a_0,a_c)$ for $+2\delta$ deviations of
$\sigma(W^+W^-)$}\label{fig15}
\end{figure}

Fig. \ref{fig15} demonstrates the considered radiative correction has
significant magnitude and its calculation increases the precision of
anomalous couplings measurement.

It must be noted that consideration of $W^+ W^- \gamma$, $ZZ\gamma$,
$Z\gamma\gamma$ processes in electron-positron annihilation  gives
additional information about $a_0$ and $a_e$, but the precision is
two orders worse  \cite{r11}. But $e^+ e^-$ beams open possibility to
measure four-boson connections \cite{r11}--\cite{r13} such as $W^+W^-$$W^+W^-$-,
$W^+W^-ZZ$-, $ZZZZ$-production that it's impossible for
$\gamma\gamma$-physics. Corresponding four-boson anomalous weak
interaction are presented by the Lagrangian with two four-dimension
operators:
\begin{equation}\begin{gathered}
L_{Q}^{(a)}=\frac{1}{4}g_{w}^{2}\left[g_{o}\left(\vec{W}_{\mu}\cdot\vec{W}^{\mu}
\right)^{2}
+\right.\\\left.+g_{c}\left(\vec{W}_{\mu}\cdot\vec{W}^{\nu}
\right)\left(\vec{W}^{\mu}\cdot\vec{W}_{\nu} \right)\right].
\end{gathered}\end{equation}
Here the first operator describes the exchange of neutral scalar
particle with very high mass, but the second one corresponds to
triplet of massive scalar particles. If four neutral boson vertex
$(ZZZZ)$ is absent (e.g. $g_0=g_c$), interaction can be realized by
massive vector boson exchange.

Using $e\gamma$ modes of two-boson production, $W^+W^-e$, $Z\gamma
e$, $ZZe$, $W^-\gamma\nu$, $W^-Z\nu$, one can consider additional
four-boson vertex $WWZ\gamma$  \cite{r14}:
\begin{equation}
\begin{gathered}
L_{n}^{(6)}=i\frac{\pi\alpha}{m_V^{2}}a_{n}\vec{W}_{\alpha}\left(\vec{W}_{\nu}\cdot\vec{W}_{\mu}^{\alpha}\right)\vec{F}^{\mu\nu}.
\end{gathered}\end{equation}
 This Lagrangian conserves $U(1)_{EM}$, $C$-,
$P$- and $SU(2)_C$-symmetry, but violates $SU(2)_L \otimes U(1)_Y$
symmetry.

From all above mentioned processes the most sensitive reactions for
$a_0$ and $a_e$ investigation are $ZZe$ and $WWe$-production. The
bounds of these constants magnitudes are one order better than in
$e^+ e^-$-process, but about 5 times worse than in
$\gamma\gamma$-mode. The vertices $\gamma\gamma\gamma Z$ and
$4\gamma$ are absent on tree level. One-loop contribution contain
both fermion loops and $W$-boson loops. The last ones give
contribution to be measured on photon collider  \cite{r15}.

\section{Conclusion}

We have analyzed the possible usage of $\gamma \gamma \to f
\bar{f} \gamma$ reaction for the luminosity measurement at
$J\!=\!0$ beams on linear photon collider. The achievable
precision of the luminosity measuring is considered and
calculated. The optimal conditions for that measurement are found
(for the high magnitude of $J\!=\!0$ cross section and small
$J\!=\!2$ background). The first-order QED correction to
$\gamma\gamma\to l \bar{l}$ cross section is calculated and analyzed
at $J\!=\!2$-beams.

The considered process gives the excellent opportunity for luminosity measurements with substantial accuracy.

The investigation of the sensitivity of process of
$\gamma\gamma\to W^+ W^-$ and $\gamma\gamma\to W^+ W^- Z$ to
genuine anomalous quartic couplings $a_0$, $a_c$ and $\tilde{a}_0$
was performed at center-of-mass energy $\sqrt{s}=1\, TeV$. It was
discovered that two-boson production has great sensitivity to
anomalous constants
$a_c$ and $a_0$ but process $\gamma\gamma\to W^+ W^- Z$ is more
suitable for study of $\tilde{a}_0$.

The fact that the minimum of the curves are close to the SM point $a_i=0$ demonstrates
the small value of the
anomalous and the standard part interference.
The first-order radiative correction to
cross section $\sigma (\gamma\gamma\to W^+ W^-)$ has significant
magnitude and its calculation increases the precision
of the $a_0$ and $a_c$ measurement.

The theoretical analysis demonstrates that investigation of
four-boson anomalous weak interaction in frame of four-dimension
anomalous Lagrangian of $\gamma\gamma$ scattering as well as in
frame of $e \gamma$ modes of two-boson production have great
importance for reconstruction gauge group of electroweak interaction
beyond the Standard theory of electroweak interaction.



\end{document}